\documentclass[a4paper,review,12pt,authoryear]{elsarticle}
\usepackage{amsmath,bm,bbm,hyperref,url,fullpage,mathtools,booktabs}
\usepackage{paralist,enumitem,todonotes,natbib}
\usepackage[toc,page]{appendix} 
\usepackage[a4paper, margin=1in]{geometry}
\usepackage{multirow}
\usepackage{algorithm}
\usepackage{algorithmicx}
\usepackage{algpseudocode}
\usepackage[labelsep=period]{caption}
\usepackage{verbatim}
\usepackage{graphicx}
\usepackage{placeins}
\usepackage{adjustbox}
\usepackage{amsfonts}
\usepackage{booktabs}
\usepackage{siunitx}
\usepackage[usestackEOL]{stackengine}

\providecommand{\tightlist}{%
  \setlength{\itemsep}{0pt}\setlength{\parskip}{0pt}}
  
\usepackage{hyperref}
\usepackage{float}
\hypersetup{
  colorlinks=false,
  linkcolor=blue,
  citecolor=blue,
  urlcolor=blue}

\let\code=\texttt
\let\proglang=\textsf
\newcommand{\pkg}[1]{{\fontseries{b}\selectfont #1}}
\usepackage{todonotes}

\journal{}
\begin{document}
\date{}
\begin{frontmatter}

  \title{Improving forecasting by subsampling seasonal time series}

  \author[mainaddress]{Xixi Li} \ead{xixi.li@manchester.ac.uk} \ead[url]{https://orcid.org/0000-0001-5846-3460}
\author[secondaryaddress]{Fotios Petropoulos} \ead{f.petropoulos@bath.ac.uk} \ead[url]{https://orcid.org/0000-0003-3039-4955}
\author[thirdaddress]{Yanfei Kang\corref{cor}} \ead{yanfeikang@buaa.edu.cn}\ead[url]{https://orcid.org/0000-0001-8769-6650}
\cortext[cor]{Corresponding author}
\address[mainaddress]{Department of Mathematics, The University of Manchester, UK.}
\address[secondaryaddress]{School of Management, University of Bath, UK.}
\address[thirdaddress]{School of Economics and Management,
	Beihang University, China.}

  \begin{abstract}
Time series forecasting plays an increasingly important role in modern business decisions. In today's data-rich environment, people often aim to choose the optimal forecasting model for their data. However, identifying the optimal model requires professional knowledge and experience, making accurate forecasting a challenging task. To mitigate the importance of model selection, we propose a simple and reliable algorithm to improve the forecasting performance. Specifically, we construct multiple time series with different sub-seasons from the original time series. These derived series highlight different sub-seasonal patterns of the original series, making it possible for the forecasting methods to capture diverse patterns and components of the data. Subsequently, we produce forecasts for these multiple series separately with classical statistical models (ETS or ARIMA). Finally, the forecasts are combined. We evaluate our approach on widely-used forecasting competition data sets (M1, M3, and M4) in terms of both point forecasts and prediction intervals. We observe performance improvements compared with the benchmarks. Our approach is particularly suitable and robust for the data with higher frequency. To demonstrate the practical value of our proposition, we showcase the performance improvements from our approach on hourly load data that exhibit multiple seasonal patterns.

  \end{abstract}


\end{frontmatter}

\newpage
\section{Introduction}
\label{introduction}

Time series forecasting is an indispensable part of modern businesses and a valuable input in the decision process for production, finance, planning, scheduling, and other activities \citep{Petropoulos2020-ec}. We are living in a big data era where large collections of time series constantly emerge. For example, large retailers need to forecast the sales for tens of thousands of products in each forecast cycle. There is a need to have robust solutions to process and forecast these large volume of series in a batch, automatic manner. However, given a set of candidate forecasting models, identifying the most appropriate and reliable model for each data is not always a straightforward task.

One way to deal with modelling time series in an automatic way would be to resort to statistical model selection approaches, like information criteria or cross-validation. Such approaches are able to select the most ``appropriate'' forecasting model, based on the in-sample fit (penalised for model complexity) or based on past performance on a hold-out set of observations. However, real-life time series data do not abide by particular data generation processes, and established patterns may change over time. In a famous quote, George Box pointed out that ``all models are wrong, but some are useful'' \citep[][ p. 424]{box1987empirical}. \citet{wolpert1996lack} also argued that a single model cannot fit in all situations. Instead, combining the forecasts across multiple models has shown not only to perform well but also to reduce the variance of the forecasts \citep{Bates1969,hibon2005combine}.

The good performance of forecast combinations is also linked with the three inherent uncertainties that a forecaster needs to tackle \citep{hyndman2008forecasting,petropoulos2018exploring}. First, the model uncertainty refers to being able to identify the correct model form. Within the exponential smoothing (ExponenTial Smoothing or Error, Trend, Seasonality; ETS) family of models, this means identifying if trend and/or seasonal components need to be included, but also their interactions (additive or multiplicative). In the autoregressive integrated moving average (ARIMA) family, tackling model uncertainty entails correctly identifying the appropriate (non-seasonal and seasonal) autoregressive and moving average terms. Second, even if the correct model form has been selected, there is uncertainty related to the estimates of the model's parameters, such as the smoothing parameters of the exponential smoothing models. Finally, data uncertainty refers to the variation of the inherent random component of the series.

While fitting many different models on the same data and averaging their forecasts is sometimes sufficient to achieve performance improvements \citep{Kolassa2011-af,Petropoulos2020-uf}, a better approach is to manipulate the data and extract additional information from them, if there is any. For example, \cite{Assimakopoulos2000-hj} and \cite{Fiorucci2016-ak}, among others, changed the local curvatures of the seasonally-adjusted data to be able to predict short and long-term behaviours. The resulting method, Theta, performed well in a variety of real-life settings, including a first place in the M3-competition \citep{Makridakis2000-oq}. \cite{Kourentzes2014-jq} and \cite{Athanasopoulos2017-ta} worked with multiple temporal aggregation levels of the same signal and performed forecast combinations across different sampling frequencies. \cite{Bergmeir2016-su} and \cite{petropoulos2018exploring} performed bootstrapping on the remainder of the time series decomposition to create several instances of the same series, which were then modelled independently and their forecasts were aggregated.

Our work builds on previous studies in an attempt to extract as much information from the data as possible. We particularly focus on the challenge of correctly modelling seasonal patterns. A seasonal time series pattern exists when the data is influenced by seasonal factors \citep[e.g., the quarter or the month of the year,][]{Rob:2011}. Seasonality is observed in
a fixed and known period, unlike other cyclical patterns. Most of the existing time series models are designed to adapt to simple seasonal patterns with short periodic cycles with respect to the series' time granularity. Examples of such simple models are the Seasonal Na\"ive method, Holt-Winters' Exponential Smoothing
\citep{winters1960forecasting}, Damped Trend Seasonal Method \citep{gardner2006exponential}, and Seasonal ARIMA (SARIMA) models. In the case of multiple, complex seasonal patterns, more complicated forecasting methods need to be considered \citep{de2011forecasting}.

For series that may display seasonal patterns, it is reasonable and feasible to use the observations from one season of the historical data to forecast the corresponding season in the future. Intuitively and similarly, several adjacent seasons could be predicted using the respective past observations. Starting with the conjecture that we can use the observations of some adjacent seasons in the past to predict the values of the corresponding seasons in the future, we construct multiple time series which consist of the observations of only one or some adjacent seasons. We refer to these new subsampled series as \textit{sub-seasonal} series, as they contain part of but not all the possible seasonal information. These derived series highlight different sub-seasonal patterns of the original time series, thus simplifying their modelling and estimation.

After constructing all possible sub-seasonal series, we extrapolate these derived series as well as the original series using standard forecasting workhorses. In this paper, we focus our empirical investigation on two widely-used forecasting families: ETS and ARIMA~\citep{Hyndman2018}. However, our framework is model-agnostic and could be expanded to any other forecasting model suitable to deal with simple seasonal patterns. In modelling each sub-seasonal series, we select the `optimal' model form and set of parameters separately and independently from other sub-seasonal series (or the original series). The forecasts produced for the different series are then combined, effectively tackling issues surrounding model and parameter uncertainty.

The key innovations of our approach are as follows:
\begin{itemize}
 \tightlist
\item We zoom in the sub-seasonal patterns of the original series that are simpler to model.
\item We do not rely on a single model and its forecasts that are based on the original series. Instead, we mitigate the importance of model selection by combining forecasts across many sub-seasonal series.
\item Our proposed approach is simple, transparent, and does not rely on a particular family of forecasting models. In particular to the latter, we do not propose a new forecasting model per se, but a framework that can be plugged into any existing model.
\end{itemize}

We conduct an extensive empirical evaluation of our proposed framework using 75 thousand real time series from the Makridakis forecasting competitions \citep{Makridakis1982-co,Makridakis2000-oq,Makridakis2020-mm}. Our framework works as a self-improving mechanism for ETS and ARIMA families of models, resulting in better point-forecast accuracy and uncertainty estimation (captured through prediction intervals) than simply applying ETS and ARIMA on the original series. We observe that improvements in performance are greater for longer forecasting horizons, where uncertainty is also higher. When applied to the case of load forecasting, our approach produces reliable forecasts, showcasing its efficacy on time series data with multiple and complex seasonal patterns.

The rest of the article is organised as follows: Section~\ref{literature} offers a short literature review on relevant studies. Section~\ref{method} describes the methodology for the proposed forecasting approach. Section~\ref{experiments} presents the experimental results and their statistical significance.
Section~\ref{application} shows the application to electric load forecasting.
Section~\ref{discussion} offers our discussions and insights. Finally, Section~\ref{conclusion} provides our conclusions.

\section{Background research}
\label{literature}

Given the plethora of available models to choose from, several selection strategies have been developed over the years. Such strategies compare the performance of different candidate models in order to choose the best one, based on some criteria. Some approaches, namely information criteria, select the model with the maximum likelihood by adding a penalty to compensate for the over-fitting of more complex models \citep{bishop2006pattern}. Popular information criteria include Akaike's Information Criterion (AIC) and Bayesian Information Criterion (BIC). However, \citet{bishop2006pattern} pointed out that information criteria could not properly account for the uncertainty of the models' parameters and tend to prefer simple models to more complex ones.

An alternative to selecting with information criteria is judging the models based on their past performance over multiple lead times. Such approaches are known as validation and cross-validation for time series, and are closely related to the concepts of fixed and rolling origin evaluation \citep{Tashman2000-mv}. \cite{Fildes2015-jo} claimed that (cross-)validation approaches are better than information criteria as the former ones take into account longer forecast horizons when evaluating past forecasts while the latter methods are based on the one-step-ahead in-sample forecast errors. \cite{montero2020fforma} achieved a good performance on the M4 forecasting competition using an approach that is based on a validation set-up enhanced by time series features. On the other hand, \cite{Billah2006-jg} showed that the performance of validation is similar to that of information criteria. In any case, (cross-)validation approaches require longer series and more computational resources.

Other approaches to model selection take a more data-driven approach, essentially attempting to answer the question: Which is the best forecasting method for my data? Many attempts have been made to tackle the task of model selection using time series features. For example, \citet{reid1972comparison} pointed out that the nature of the data has an influence on the performance of the forecasting methods. \citet{collopy1992rule} developed a rule-based system containing 99 rules to produce forecasts based on the features of the data. \citet{adya2001automatic} identified time series features automatically for rule-based forecasting. \citet{petropoulos2014horses} explored the factors that affect forecasting accuracy in the field of demand forecasting, and proposed related selection protocols. \citet{kang2017visualising} used Principal Component Analysis (PCA) to visualise the forecasting algorithm performance in the time series instance spaces and had a better understanding of their relative performance. Finally, \citet{talagala2018meta} used a decision tree to select the best forecasting method based on 42 manual features.

Regardless of the available options to perform model selection between forecasting models, the \textit{No Free Lunch} theorem suggests that there does not exist one method that will perform always best across series, nor across time, due to the dynamic nature of the problem. As such, instead of considering selecting a single model (per series or across series), combinations of forecasts are an alternative way forward. Forecast combinations tend to yield better results when one is averaging forecasts from robust models but also when there is significant diversity among the forecasts \citep[]{wang2016select,thomson2019combining,Lichtendahl2020-bj,KANG2021}. Combinations of forecasts address two of the sources of uncertainty in forecasting \citep{petropoulos2018exploring}, model's form and model's parameters uncertainty, as they do not rely on a single model anymore.

Apart from obtaining diverse and robust forecasts, another critical factor in the efficiency of forecast combinations is the estimation of the weights. Recently scholars tried to use advanced machine learning techniques to optimise optimal weights with a focus on the time series features of the target time series. \citet{montero2020fforma} trained XGBoost \citep{Chen2016XGBoost} to obtain optimal weights for each method based on the 42 manual time series features, achieving the second-best performance in the M4 Competition. \citet{li2020forecasting} used a similar approach to estimate optimal combination weights on the automatic image features of the series and produced comparable performance with the top performers in M4 Competition. Regardless of the good performance of some optimal-weighting combination strategies, equal-weights combinations also perform remarkably well \citep{Petropoulos2020-uf}, and often are hard to beat by other more complex approaches \citep{Jose2008-pa,Genre2013-ka}. This is referred to as the ``forecast combination puzzle'' in the literature~\citep{Watson2004Combination,smith2009simple,claeskens2016forecast}.

A special case in forecast combinations refers to approaches that, instead of fitting different models to the same data, manipulate the original data and create other series in an attempt to amplify some particular series characteristics while suppressing others. One such approach can be found in the core of the theta method \citep{Assimakopoulos2000-hj}. Instead of applying forecasting models on the original data, the theta method works on the seasonally-adjusted data. These are then further decomposed into theta lines, aiming to capture the short and long-term dynamics. The theta method essentially changes the data by manipulating the residual of the regression model. This simple method is a robust benchmark in the forecasting literature and was the top-performing method at the M3 forecasting competition \citep{Makridakis2000-oq}.

Another approach that also uses data manipulation to extract additional information is (non-overlapping) temporal aggregation \citep{Nikolopoulos2011-bt}. This technique down-samples the original series in order to obtain more series of lower frequencies. Higher temporal aggregation levels offer smoother, less intermittent series, allowing for better extrapolation of trend patterns. Still, lower levels of aggregation allow for better estimation of seasonal patterns. Producing forecasts for many levels independently and then combining such forecasts has shown sizable improvements in accuracy both for fast and slow-moving series \citep{Kourentzes2014-jq,Petropoulos2015-vq}. More recently, the combination of forecasts from several temporal aggregation levels is made in a hierarchical fashion \citep{Athanasopoulos2017-ta,kourentzes2021elucidate}.

A third approach that follows the concept of extracting more information from the data is bootstrapping~\citep[e.g.,][]{Bergmeir2016-su,hasni2019spare}. In particular for time series forecasting, \cite{Bergmeir2016-su} proposed decomposing the original signal to the trend, seasonal, and remainder components, bootstrapping the latter in order to create several series of remainders, and reconstructing the series by putting together trend, seasonality, and each bootstrapped series of remainders. Following the above process, one can construct several series of the same structure (trend and seasonality) but with different noise. Instead of simply forecasting the original series, one can forecast all reconstructed series and then aggregate (average) the forecasts. `Bagging', when applied in conjunction with an algorithm that considers a large number of candidate models and automatically selects the best one (such as automatic exponential smoothing or ARIMA), can tackle the three sources of uncertainty: model form, model parameters, and data \citep{petropoulos2018exploring}.

\section{Forecasting through integrating the forecasts of multiple time series with different sub-seasonal patterns}
\label{method}

\begin{figure}[ht!]
	\centering
	\includegraphics[width=1\linewidth]{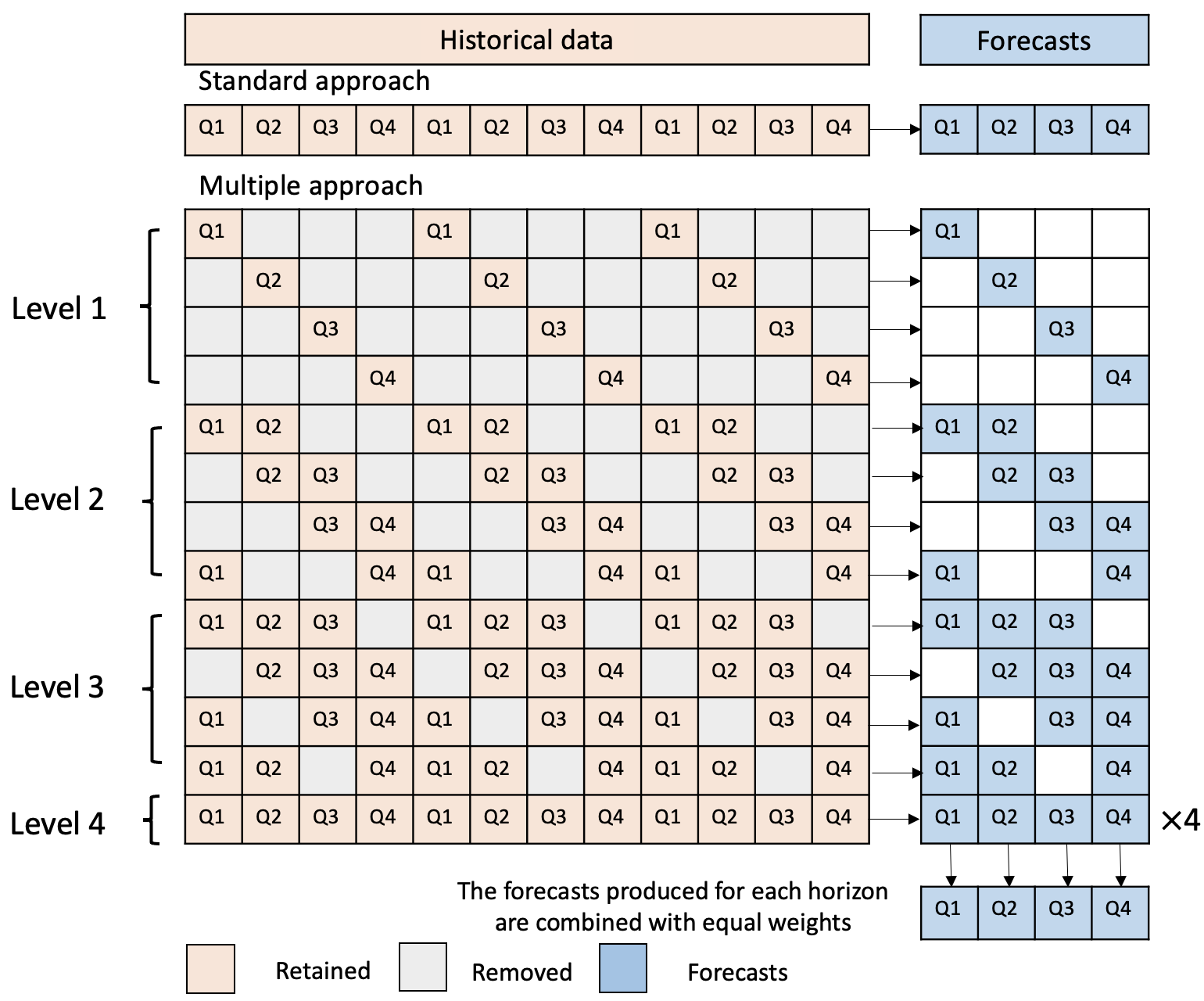}
	\caption{
	Caption: A graphical illustration of the standard approach and the proposed multiple approach applied to a quarterly series. \\
	Figure~\ref{fig:framework}. Alt text: Sequences of adjacent boxes showing the standard approach and the steps of multiple approach applied to a quarterly series, with each box representing one observation.\\
	Figure~\ref{fig:framework}. Long description: A graphical illustration of the proposed approach applied to a quarterly series with three years of historical observations. The aim is to forecast the next year (four quarters). The light orange, grey and blue boxes represent the retained, the removed  observations and forecasts respectively. The forecasting procedure consists of three main steps: 1) create new time series only containing one or more adjacent seasons, i.e., $Q_1$, $Q_2$, $Q_3$, $Q_4$, $Q_1\&Q_2$, $Q_2\&Q_3$, $Q_3\&Q_4$, $Q_4\&Q_1$, $Q_1\&Q_2\&Q_3$, $Q_2\&Q_3\&Q_4$, $Q_3\&Q_4\&Q_1$, $Q_4\&Q_1\&Q_2$, and the original series $Q_1\&Q_2\&Q_3\&Q_4$. The level value in the figure represents the corresponding several adjacent quarterly observations are retained, 2) forecast separately each sub-seasonal and the original series, and 3) combine the produced forecasts with equal weights. Note that when combining, $Q_1\&Q_2\&Q_3\&Q_4$ is repeated four times so that equal importance is given to all information levels.}
	\label{fig:framework}
\end{figure}

Suppose we observe a quarterly series of three years, and our target is to forecast the next year (four quarters). The standard approach would be to consider all 12 available data points and forecast using a time series method. In our approach, we construct various sub-seasonal series (e.g., series consisting of only some of the quarters of each year) from the original target series. The final forecasts can be obtained by combining the forecasts produced from the series with different sub-series patterns. Figure~\ref{fig:framework} graphically illustrates our method. For this particular series, the proposed forecasting procedure is as follows.

  \begin{enumerate}
\def\labelenumi{\arabic{enumi}.}
\tightlist
  \item Create a new series that consists only of the first quarters ($Q_1$) of each year. Repeat for the  second ($Q_2$), third ($Q_3$), and fourth ($Q_4$) quarters of the year. This is the first level of information, where each sub-seasonal series contains exactly one quarter.
  \item Create a new series that only consists of two adjacent quarters of each year. Specifically, extract the observations of the first and second quarters ($Q_1\&Q_2$) of each year and construct a new series accordingly. Repeat for $Q_2\&Q_3$, $Q_3\&Q_4$ and $Q_4\&Q_1$ (or $Q_1\&Q_4$). This is the second level of information, where each sub-seasonal series contains exactly two adjacent quarters.
  \item Create a new series that only consists of three adjacent quarters of each year. Specifically, extract the observations of the first, second and third quarters ($Q_1\&Q_2\&Q_3$) of each year and construct a new series accordingly. Repeat for $Q_2\&Q_3\&Q_4$, $Q_3\&Q_4\&Q_1$ (or $Q_1\&Q_3\&Q_4$) and $Q_4\&Q_1\&Q_2$ (or $Q_1\&Q_2\&Q_4$). This is the third level of information, where each sub-seasonal series contains exactly three adjacent quarters.
  \item Forecast the constructed sub-seasonal series, and the original series ($Q_1\&Q_2\&Q_3\&Q_4$, which is the fourth level of information), using standard time series forecasting methods (e.g., ETS or ARIMA). That is, we use the historical observations of some adjacent seasons to forecast the corresponding seasons in the future. 
  \item Combine the obtained forecasts with equal weights. Note that the original series, $Q_1\&Q_2\&Q_3\&Q_4$, is repeated four times so that equal importance is given to all information levels.
\end{enumerate}

More generally, assume we are interested in forecasting a seasonal time series $\{y_t, t = 1, 2, \cdots, T\}$ with a sampling frequency $m$, the forecasting procedure through integrating the forecasts of multiple time series with different sub-seasonal patterns is as follows.

\begin{enumerate}
\def\labelenumi{\arabic{enumi}.}
\tightlist
	\item  \textbf{Constructing} multiple series with sub-seasonal patterns. Create new time series that contain observations of only one or a few adjacent seasons and the	corresponding frequency is equal to the number of seasons. 
	
	\item  \textbf{Forecasting} each of the newly derived series using standard time series forecasting methods. Theoretically, the number of the constructed series is $m^2$ for all the information levels. However, in practice, we only need to forecast the original series once for the last information level, and when the forecasting horizon $h$ is less than the frequency $m$, we do not need to construct and forecast all the subsampled series. Therefore, the number of subsampled series $M$ that need to be forecast can be written as
\[ M =
  \begin{cases*}
  \frac{(m-h)(m+h-1)}{2}+(h-1)m+1, & if $h < m$, \\
    m(m-1)+1, & otherwise. \\
  \end{cases*}\]

	\item  \textbf{Combining} the forecasts produced from the sub-seasonal series with equal weights. Note that when combining, the original series in the last information level is repeated $m$ times so that equal importance is given to all information levels. Since the `optimal' model form and set of parameters are estimated separately and independently when modelling each sub-seasonal series, forecast combination makes it possible to tackle issues around model and parameter uncertainty~\citep{petropoulos2018exploring}.
\end{enumerate}
In the following sections, we introduce each step of our proposed method in detail.
\FloatBarrier

\subsection{Constructing multiple time series with sub-seasonal patterns}\label{sec:construction}

\begin{figure}[!ht]
	\centering
	\includegraphics[width=1\linewidth]{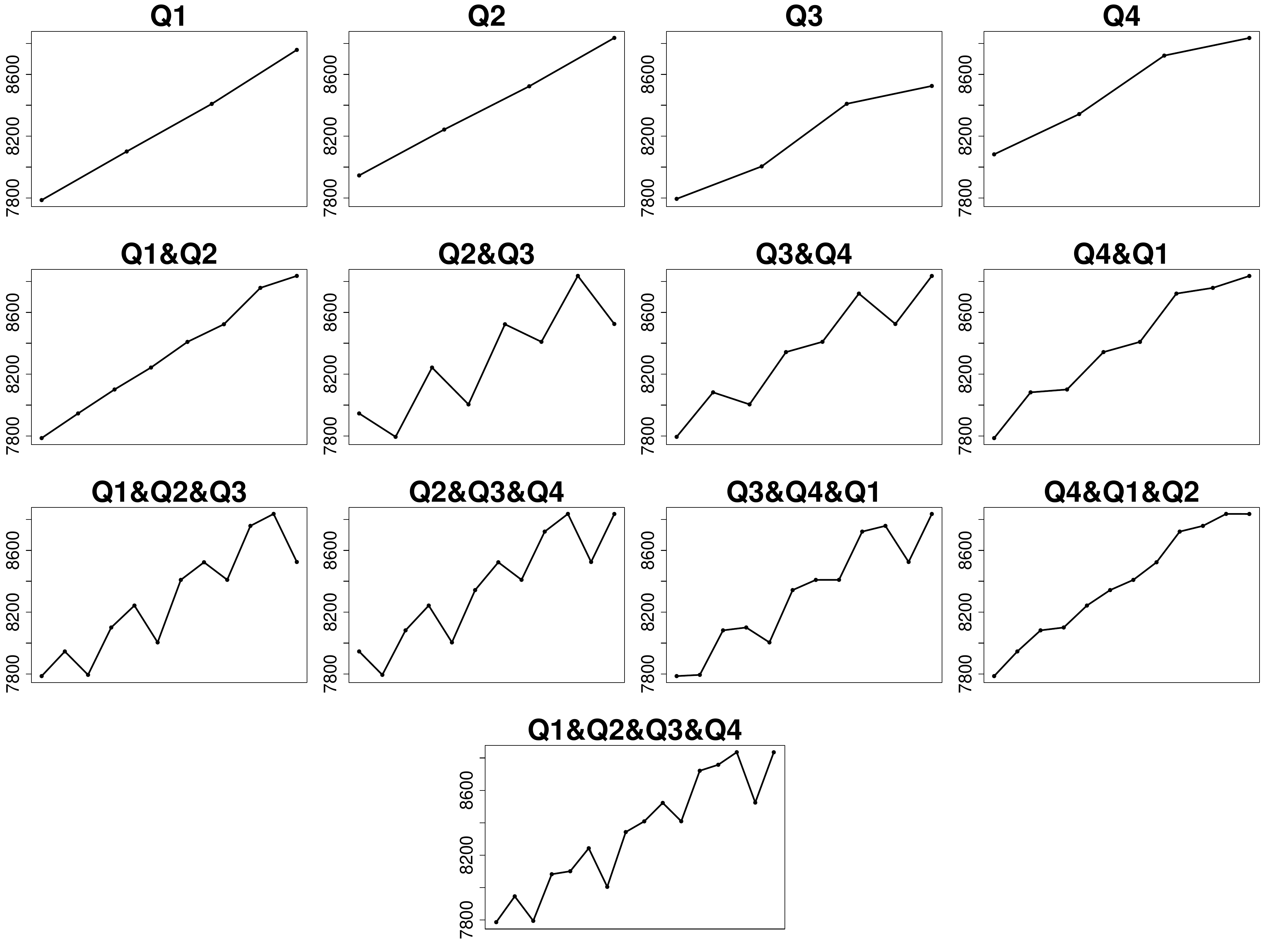}
	\caption{Caption: An example visualising the constructed time series by our proposed method. The quarterly time series Q520 from M3 competition is used to be the target series. Thirteen sub-seasonal series that contain the observations of only one or a few adjacent quarters are constructed.\\
	Figure~\ref{fig:Q520}. Alt Text: Thirteen line graphs in four rows showing the constructed sub-seasonal series that contain the observations of only one or a few adjacent quarters of an example series Q520 from M3 competition.}
	\label{fig:Q520}
\end{figure}

In this section, we use the quarterly time series Q520 from the M3 competition data set~\citep{Makridakis2000-oq} to demonstrate the construction of sub-seasonal series from the original time series and illustrate the importance of forecasting with sub-seasonal patterns. Figure~\ref{fig:Q520} shows the constructed sub-seasonal series. For this particular time series, we can construct $m(m-1)+1=13$ series with different sub-seasonal patterns containing the observations of only one or a few adjacent quarters. For each sub-seasonal series, the corresponding frequency is equal to the number of the adjacent quarters. Note that among the thirteen series, the last series shown in the last row of Figure~\ref{fig:Q520} corresponds to the original, target series. The series in each row of Figure~\ref{fig:Q520} contain the same number of quarters, i.e., belong to the same information level. In particular, each series in the first row contains observations of only one quarter from the original series. The series in the second row contain observations of two adjacent quarters, and the third row shows the sub-seasonal patterns from observations of three adjacent quarters.

We observe from Figure~\ref{fig:Q520} that each newly created series highlights different sub-seasonal patterns and components of the original series. 
The four series in the first row, containing only one quarter, show an upward trend, while the trend in series that contains $Q_1$ or $Q_2$ is almost linear. From the second row of Figure~\ref{fig:Q520}, we can see that the sub-seasonal series containing $Q_1\&Q_2$, $Q_2\&Q_3$, $Q_3\&Q_4$ and $Q_4\&Q_1$ exhibit different amounts of variability.  Also, all four series demonstrate an increasing trend. 
The first three series in the third row demonstrate larger variability around the upward trend. However, these patterns differ across the series. Moreover, the ranges of the values of the observations from different sub-seasonal patterns differ. 
Thus, the constructed series highlight different sub-seasonal patterns of the target series, which can be used to improve the forecasting performance.

\subsection{Producing forecasts for each sub-seasonal series}

After constructing multiple sub-seasonal series,  we apply two families of models, namely ETS and ARIMA, on each of the constructed series to forecast the corresponding seasons in the future. Table~\ref{tab:two-methods} provides details of the implementations used.

ETS is able to capture the components of a time series: error, trend and seasonality \citep{hyndman2002state}. The model form can be abbreviated as ETS (error, trend, seasonality). The error component can be additive (``A'') or multiplicative (``M''). The trend and seasonality terms can be none (``N''), ``A'' or ``M'', while the trend can additionally be damped or not. The corrected Akaike's Information Criterion (AICc) is used to determine which of the ETS models is most appropriate for a given series. The ETS model selection and the corresponding parameter estimation for each sub-seasonal series are made using the function \code{ets()} in the \proglang{R} package \pkg{forecast}~\citep{Rforecast}.

\begin{table}[ht]
  \caption{The standard methods used to forecast each sub-seasonal series, which are implemented in the \pkg{forecast} package of \proglang{R}.}
\label{tab:two-methods}
\centering
\resizebox{\textwidth}{!}{
  \begin{tabular}{p{0.3\columnwidth}p{0.55\columnwidth}p{0.3\columnwidth}}
  \toprule
  Forecasting  method& Description       & \proglang{R} function         \\
  \midrule
  ETS                & The exponential smoothing state family of models \citep{hyndman2002state}.                                                                                   &\code{ets()}                  \\
  ARIMA& The autoregressive integrated moving average family of models \citep{HK08}.                                                                & \code{auto.arima()}           \\
  \bottomrule
  \end{tabular}}
\end{table}

We also consider ARIMA models  based on an algorithm that searches for the model form with the smallest AICc value and estimates the corresponding parameters \citep{Rforecast}. By default, it combines unit root tests, minimisation of the AICc, and Maximum Likelihood Estimation (MLE) to obtain the final ARIMA model among models with various AutoRegressive (AR) orders and moving average (MA) orders, up to a maximum of five.

It is worth emphasising that the constructed sub-seasonal series can also be forecast with other standard methods. Moreover, one may choose to apply different families of models on each constructed series or each information level. In any case, a different model form and set of parameters are selected for each sub-seasonal series, and a different set of forecasts is produced accordingly. The produced sets of forecasts for the sub-seasonal series are used in the next step: forecast combination.

\subsection{Forecast combination}

After extrapolating each series separately, forecasts produced from the multiple sub-seasonal series are averaged with equal weights (see Figure~\ref{fig:framework}).
In this section, in order to illustrate how our method works, we use the same example as in Section~\ref{sec:construction} to compare the combined forecasts with the standard forecasts of the target time series, based on the ETS model. 

\begin{figure}[h!]
	\centering
	\includegraphics[width=1\linewidth]{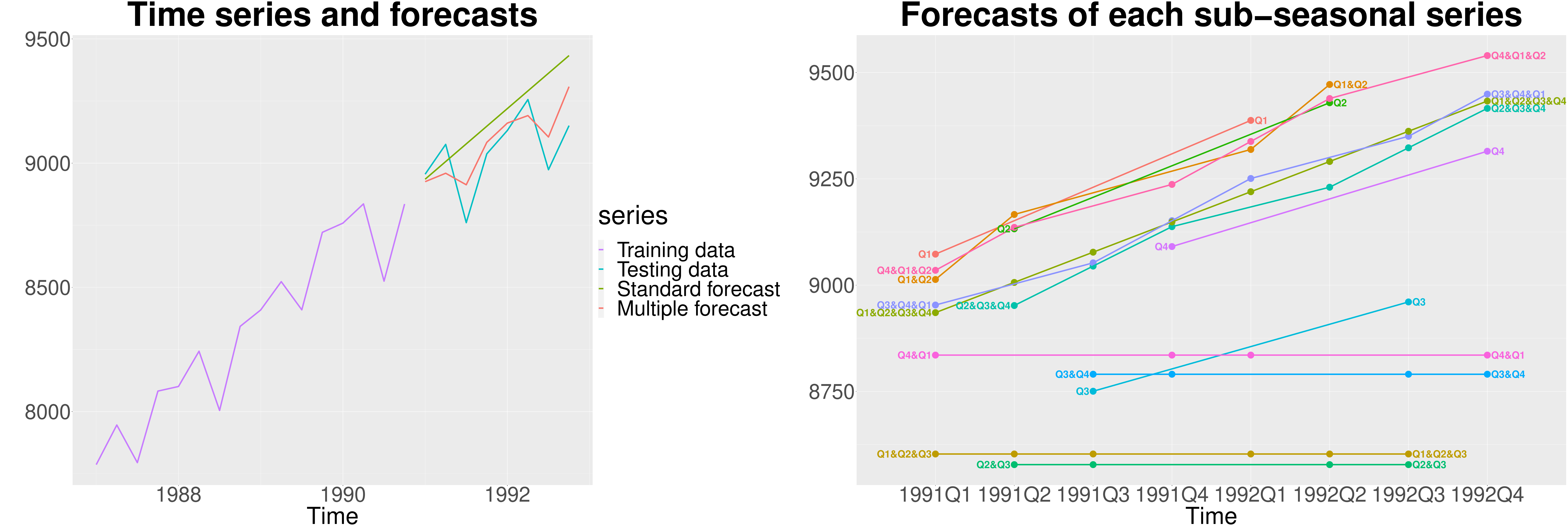}
	\caption{Caption: Left: standard and multiple forecasts of Q520 from M3 competition. Right: individual ETS forecasts for each sub-seasonal series.\\
	Figure~\ref{fig:Q520-ets}. Alt Text: Two panels, with the left line graph showing the standard and multiple forecasts of Q520 from M3 competition, and the right line graph showing individual ETS forecasts for each sub-seasonal series.}
	\label{fig:Q520-ets}
\end{figure}

Figure~\ref{fig:Q520-ets} visualises the historical and test data of the quarterly time series Q520 from the M3 competition, and their standard (using the original series only) and multiple (using all sub-seasonal series) forecasts, respectively. The right panel of Figure~\ref{fig:Q520-ets} visualises the forecasts of each sub-seasonal series individually. 
We can see that our method efficiently forecasts the variation as the information level of the series decreases (fewer adjacent quarters are considered) compared with the standard ETS method. This shows that even if the modelling for the original series ``fails'' and a seasonal model is not selected, the availability of further sub-seasonal patterns will render the forecasts more robust due to the capacity of capturing the diverse patterns. As a result, no single sub-seasonal series might be enough to offer improvements over the standard approach of modelling the original series. The benefit of our approach stems from the combination of `suboptimal' forecasts from all the sub-seasonal series. The combination (or aggregation) of forecasts from different ``interpretations'' of the original series has been successfully applied before in the contexts of multiple temporal aggregation \citep{Petropoulos2014-ad} and bagging \citep{Bergmeir2016-su}.

\begin{figure}[h!]
	\centering
	\includegraphics[width=1\linewidth]{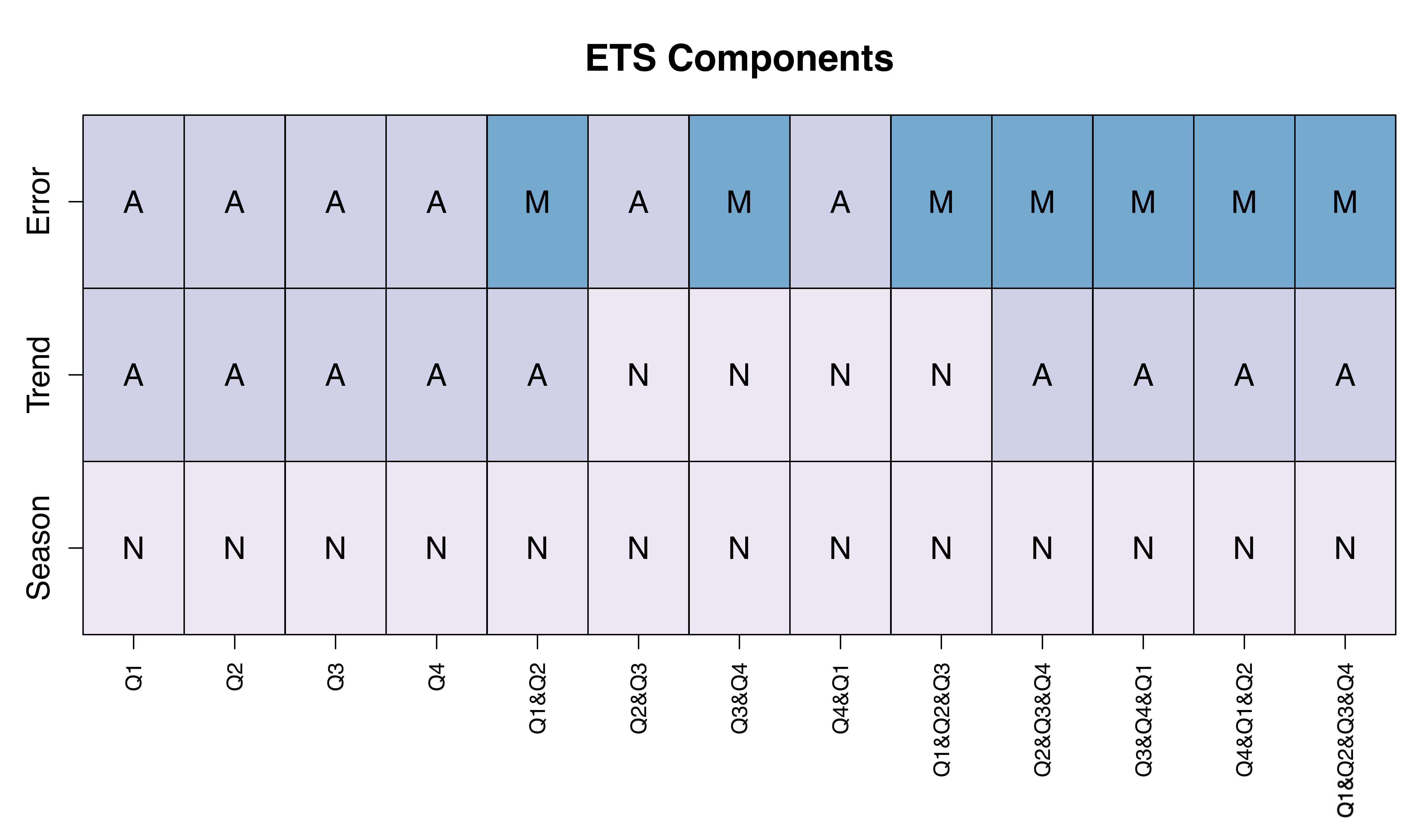}
	\caption{Caption: Optimal ETS components for each sub-seasonal series of Q520 from M3 competition data.\\
	Figure~\ref{fig:Q520-ets-components}. Alt Text: Three rows of adjacent boxes showing the optimal ETS components for each sub-seasonal series of Q520 from M3 competition data, with each box showing ``N'', ``A'' or ``M'' for the corresponding optimal component.}
	\label{fig:Q520-ets-components}
\end{figure}

The diverse patterns captured by the sub-seasonal series are presented in Figure~\ref{fig:Q520-ets-components}. Despite the fact that no sub-seasonal series is identified as seasonal, the combined forecast from our approach displays seasonality due to the proper alignment of the sub-seasonal forecasts to the corresponding periods and the different degrees of trend identified (see, for example, the right panel of Figure~\ref{fig:Q520-ets} and the $Q_1\&Q_2$ series).

We also use the proposed methodology to produce prediction intervals (PIs) by simply averaging the PIs produced from the subsampled series.  That is, for each forecasting horizon, we calculate the PI lower bound by combining the lower bounds of the PIs generated from the corresponding sub-seasonal series with equal weights. The same process applies for calculating the upper bounds.

\section{Empirical evaluation on M competitions}
\label{experiments}
\subsection{Data}
\label{sec:data}
Table~\ref{data_description} presents the data used in our empirical experiments. We consider the quarterly, monthly and hourly subsets of M1~\citep{Makridakis1982-co}, M3~\citep{Makridakis2000-oq}, and M4~\citep{Makridakis2020-mm} forecasting competitions, which refer to various categories such as demographic, industry, finance, economics, and others. The M1 and M3 data sets are publicly available in the \pkg{Mcomp} \proglang{R} package~\citep{Mcomp}, and the M4 data set in the \pkg{M4comp2018} \proglang{R} package~\citep{montero2018m4comp2018}. The forecasting horizons for quarterly, monthly, and hourly data are 8, 18, and 48, with the total number of series being 24,959, 50,045, and 414, respectively. The inclusion of the hourly data offers an evaluation of our approach on series with higher frequencies.

\begin{table}[H]
	\normalsize
	\centering
	\caption{Data used for the empirical evaluation. $h$ denotes the forecasting horizon.}
	\label{data_description}
	\scalebox{0.8}{
	\begin{tabular}{llccc}
		\midrule
		Category &Source&Number of series&Frequency& $h$ \\
		\midrule
		\multirow{3}{*}{Quarterly}	&  M1   & 203 & 4  &{8}   \\
		& M3   & 756  &4  & 8  \\
		& M4   & 24000  &4  &  8\\
		&Total&24959\\
		\midrule
		\multirow{3}{*}{Monthly} & M1   &617 &12 &{18}\\
		& M3   &1428 &12 &18\\
		& M4  &48000 &12 &18\\
		&Total&50045\\
		\midrule
		Hourly& M4  & 414  &24  &48\\
		\bottomrule
	\end{tabular}
	}
\end{table}

\subsection{Evaluation metrics}
To evaluate the accuracy of the point forecasts, we use a commonly employed measure, the Mean Absolute Scaled Error~\citep[MASE,][]{hyndman2006another}. This accuracy measure was also used in the M4 competition. It is calculated as:
\begin{equation*}
\begin{aligned}
\mathrm{MASE}&=\frac{1}{h}\frac{\sum_{t=1}^h\mid y_{T+t}-\hat{y}_{T+t} \mid}{\frac{1}{T-m}\sum_{t=m+1}^T\mid y_{t}-y_{t-m} \mid},\\
\end{aligned}
\end{equation*}
where $y_{T+t}$ is the real value of the time series at time point $T+t$, $T$ is the length of the historical data, $\hat{y}_{T+t}$ is the
point forecast, $h$ is the forecasting horizon and $m$ is the frequency of the data (e.g.,
$4$, $12$, and $24$ for quarterly, monthly, and hourly series, respectively).

In addition to MASE, the Absolute value of the Scaled Mean Error~\citep[ASME,][]{spiliotis2019improving} is employed to measure the bias of the forecasts. ASME is calculated as:
$$
\mathrm{AMSE}=\left|\frac{1}{h}\frac{\sum_{t=1}^h (y _{T+t}-\hat{y}_{T+t})}{\frac{1}{T} \sum_{t=1}^{T} y_{t}}\right|.
$$

In order to quantify the performance on forecasting uncertainty, we use the $(1-\alpha) \times 100 \%$ prediction intervals (PIs), and apply the Mean Scaled Interval Score~\citep[MSIS,][]{Gneiting2007a} to measure the accuracy of PIs.
\begin{equation*}
\resizebox{\textwidth}{!} 
{
$\text{MSIS}=\frac{1}{h} \frac{\sum_{t=1}^{h}\left(f_{T+t}^{u}-f_{T+t}^{l}\right)+\frac{2}{\alpha}\left(f_{T+t}^{l}-y_{T+t}\right) \mathbbm{1}\left\{y_{T+t}<f_{T+t}^{l}\right\}+\frac{2}{\alpha}\left(y_{T+t}-f_{T+t}^{u}\right) \mathbbm{1}\left\{y_{T+t}>f_{T+t}^{u}\right\}}{\frac{1}{T-m} \sum_{t=m+1}^{T}\left|y_{t}-y_{t-m}\right|}$,}
\end{equation*}
where $f_{T+t}^{l}$ and $f_{T+t}^{u}$ are the lower and upper bounds of the generated prediction intervals, where we set $\alpha=0.05$ (corresponding to 95\% prediction intervals), and $\mathbbm{1}$ is the indicator function, which equals to 1 when the condition is true and 0 otherwise.

\subsection{Standard versus multiple forecasts}
To evaluate the proposed method, we compare it with the standard benchmarks (i.e., ETS and ARIMA applied on the original series) with regard to the MASE, AMSE, and MSIS values for the 95\% confidence level over the quarterly, monthly, and hourly data sets.  Table~\ref{mean-accuracy} presents the forecasting performance regarding the mean MASE, AMSE, and MSIS values for short-term, medium-term, long-term, and overall horizons over the M1, M3, and M4 quarterly, monthly, and hourly subsets.  ``Standard'' represents the standard benchmarks, while ``Multiple'' represents the proposed method. We observe that:

\begin{itemize}
\tightlist
	\item For quarterly data, we find that ``Multiple'' performs better than the standard benchmarks in most mid and long-term horizons (when $h>3$) in terms of the forecasting accuracy and bias.
	
	\item For monthly data, ``Multiple'' performs very competitively against ``Standard'' for almost all the forecasting horizons regarding the mean results.  Our approach is also less biased, based on AMSE.
	
	\item For hourly data, ``Multiple'' performs better than ``Standard'' for all the forecasting horizons. The differences are very large for the case of ETS, where the drops in the values of the mean MASE and mean MSIS are more than 40\%. Also, our approach is less biased for all the horizons in terms of AMSE.
	
  \item The proposed approach is more suitable for the time series with higher frequencies (higher values of $m$), which yield a larger number of sub-seasonal series. The diversity in the sub-seasonal patterns is beneficial in improving the forecasting performance by forecast combination.

\end{itemize}

\begin{center}
\begin{table}[H]
	\normalsize
	\centering
	\caption{Benchmarking the performance of our proposed method against all the benchmark
		models with regard to the mean of MASE, AMSE and MSIS
		values for the 95\% confidence level over the M1, M3 and M4  quarterly, monthly and hourly data sets. Entries in \textbf{bold} highlight that our method outperforms the corresponding benchmarks.}
	\label{mean-accuracy}

	\begin{adjustbox}{width=1\textwidth}
		\begin{tabular}{lllllllllllllllll}
			\toprule
&&\multicolumn{5}{c} {Quarterly}&\multicolumn{5}{c} {Monthly}&\multicolumn{5}{c} {Hourly}\\
\cmidrule(lr){3-7} \cmidrule(lr){8-12} \cmidrule(lr){13-17} 

&&	\multicolumn{15}{c}{MASE}   \\
			&&$h$=1 &1-3&4-6&7-8&1-8 &$h$=1&1-6&7-12&13-18&1-18 &$h$=1&1-16&17-32&33-48&1-48 \\
			\cmidrule(lr){3-7} \cmidrule(lr){8-12}\cmidrule(lr){13-17} 
\multirow{2}{*}{ETS} 	&Standard   & 0.599   &0.774  & 1.261  &1.607 &1.165& 0.454  &0.651  & 0.965    &1.225&0.947& 0.390 &1.410  &1.611    &2.450&1.824\\
	&	Multiple &0.644 &0.797 & \textbf{1.252} & \textbf{1.567}  & \textbf{1.160}&\textbf{0.451} &\textbf{0.633} &\textbf{0.936} &\textbf{1.173}   &\textbf{0.914} &\textbf{0.375}&\textbf{0.879}&\textbf{0.919}
	&\textbf{1.303}&\textbf{1.034}\\
				\cmidrule(lr){3-7} \cmidrule(lr){8-12}\cmidrule(lr){13-17} 
	
\multirow{2}{*}{ARIMA} 	&Standard         & 0.596&0.779 & 1.273  &1.605&1.171 & 0.441 &0.627 & 0.953  &1.213&0.931& 0.366& 0.708& 0.867& 1.272& 0.949  \\

	&	Multiple  &0.641  &{0.807} & {1.274} &\textbf{1.584} &{1.177}  &0.442 & 0.628 &\textbf{0.932} & \textbf{1.172}&\textbf{0.911}&\textbf{0.336}
	&\textbf{0.648}&\textbf{0.799}&\textbf{1.157}&\textbf{0.868}\\
			\cmidrule(lr){3-7} \cmidrule(lr){8-12}\cmidrule(lr){13-17} 

			&&	\multicolumn{15}{c}{AMSE}   \\
			
			&&$h$=1 &1-3&4-6&7-8&1-8 &$h$=1&1-6&7-12&13-18&1-18&$h$=1&1-16&17-32&33-48&1-48 \\
			\cmidrule(lr){3-7} \cmidrule(lr){8-12} \cmidrule(lr){13-17}
\multirow{2}{*}{ETS} 	&Standard&0.086& 0.091& 0.149& 0.194& 0.126&0.079& 0.084& 0.128& 0.168&0.117&0.038& 0.179& 0.136& 0.239& 0.172  \\
	&	Multiple &0.089& 0.092&\textbf{0.146}&\textbf{0.188}&\textbf{0.123}&\textbf{0.078}&\textbf{0.079}& \textbf{0.122}&\textbf{0.156}&\textbf{0.108}&\textbf{0.036}&\textbf{0.110}& \textbf{0.085}&\textbf{0.135}&\textbf{0.100}     \\
			\cmidrule(lr){3-7} \cmidrule(lr){8-12}  \cmidrule(lr){13-17}
\multirow{2}{*}{ARIMA} 	&Standard &0.083& 0.091& 0.150& 0.195& 0.127&0.076& 0.079& 0.125& 0.164& 0.113&0.043& 0.076& 0.080& 0.122& 0.080     \\
	&	Multiple &0.088& 0.094&\textbf{0.149}&\textbf{0.191}&\textbf{0.126}&0.076& \textbf{0.078}&\textbf{0.121}&\textbf{0.155}&\textbf{0.108}&\textbf{0.035}&\textbf{0.072}&\textbf{0.074}&\textbf{0.114}&\textbf{0.075}     \\
			\cmidrule(lr){3-7} \cmidrule(lr){8-12}  \cmidrule(lr){13-17}

&&	\multicolumn{15}{c}{MSIS}   \\
			
		&&$h$=1 &1-3&4-6&7-8&1-8 &$h$=1&1-6&7-12&13-18&1-18&$h$=1&1-16&17-32&33-48&1-48  \\
			\cmidrule(lr){3-7} \cmidrule(lr){8-12} \cmidrule(lr){13-17} 
\multirow{2}{*}{ETS} 	&Standard      & 4.729  &6.154   &10.354 &13.585 &9.587& 3.698&5.137   & 8.493    &11.143 &8.258& 3.127&11.685   & 15.107    &25.669 &17.487 \\
	&	Multiple  &{5.046}  &{6.253}  &\textbf{10.004} &\textbf{12.908}&\textbf{9.323}  &\textbf{3.645}  &\textbf{5.023} &\textbf{8.109}&\textbf{10.409}&\textbf{7.847}&\textbf{2.923}&\textbf{6.534}&\textbf{8.937}&\textbf{12.999}&\textbf{9.490} \\
		\cmidrule(lr){3-7} \cmidrule(lr){8-12}\cmidrule(lr){13-17} 
\multirow{2}{*}{ARIMA} 	&Standard       &5.543&7.221    &12.230   &15.787&11.241 & 4.045&5.470   &9.132   &11.640 &8.747&3.295& 5.753& 7.463& 9.279& 7.498  \\

	&	Multiple&\textbf{5.480}  &\textbf{7.159}&\textbf{11.848}&\textbf{15.176}&\textbf{10.921}&\textbf{3.936}  &\textbf{5.458} &\textbf{8.909} &\textbf{11.119 }&\textbf{8.495} &\textbf{3.141}&\textbf{5.248}&\textbf{6.533}&\textbf{7.889}&\textbf{6.557} \\
			
			\midrule
		\end{tabular}
	\end{adjustbox}
\end{table}
\end{center}

\subsection{Significance tests}

To investigate the statistical significance of the performance differences between the proposed method and the benchmarks, we carry out Diebold-Mariano (DM) tests~\citep{harvey1997testing}. In DM tests, the null hypothesis is that the two approaches have the same forecast accuracy. We report the percentage of times that the DM tests statistic falls in the lower or upper 2.5\%  tail of a standard Normal distribution. The DM tests is implemented using \code{forecast::dm.test()} in \proglang{R}. The entries in Table~\ref{dm-test}  present the percentage of times ``Multiple'' is significantly better or worse than the standard benchmarks for different horizons. We observe that the percentage of times that our approach outperforms the benchmarks significantly increases as the horizon and frequency increase. For instance, the multiple ETS performs significantly better than the standard ETS in 53\% of the cases for the longer horizons (33-48 hours ahead) of the hourly data, and only 18\% of the cases worse.

\begin{center}
\begin{table}[H]
	\normalsize
	\centering
	\caption{Diebold-Mariano (DM) tests for comparing the forecasting accuracy of the standard method with the multiple method.
	The entries show the percentage of times ``Multiple'' is significantly better or worse than ``Standard'' for different horizons.}
	\label{dm-test}

	\begin{adjustbox}{width=1\textwidth}
		\begin{tabular}{llllllllllllll}
			\toprule

&&\multicolumn{4}{c} {Quarterly}&\multicolumn{4}{c} {Monthly}&\multicolumn{4}{c} {Hourly}\\
\cmidrule(lr){3-6} \cmidrule(lr){7-10} \cmidrule(lr){11-14} 
	&&$h$=1-3 &4-6&7-8&1-8&$h$=1-6&7-12&13-18&1-18&$h$=1-16&17-32&33-48&1-48 \\
			\cmidrule(lr){3-6} \cmidrule(lr){7-10}\cmidrule(lr){11-14} 

\multirow{2}{*}{Multiple ETS}  &better & 2.893&\textbf{3.698}&\textbf{4.559}&15.313 &\textbf{16.275}&\textbf{19.085}&\textbf{29.190}&\textbf{33.276} &\textbf{53.382}&\textbf{44.686}&\textbf{53.382}&\textbf{61.353}  \\
	 &worse&6.431&2.821&3.362&18.611&16.040&14.655&23.039&24.464 &11.836&10.870&18.116&14.010\\
\cmidrule(lr){3-6} \cmidrule(lr){7-10}\cmidrule(lr){11-14}
	
\multirow{2}{*}{Multiple ARIMA} & better&2.757&\textbf{3.666}&\textbf{3.971}&13.566&14.015&\textbf{20.054}&\textbf{28.045}&\textbf{30.884 }&\textbf{31.643}&\textbf{32.367}&\textbf{44.444}&\textbf{51.691} \\

	 &worse &5.862&3.337&3.277&17.749&16.759&15.374&22.476&24.886&10.870&13.285&13.043&13.768  \\

			\midrule
		\end{tabular}
	\end{adjustbox}
\end{table}
\end{center}

\subsection{Linking time series features to the performance of the proposed approach}

To gain a better understanding of how the time series features affect the performance of the proposed approach and further investigate when our method works well, we consider dividing our data into four categories: series with no trend nor seasonality, series with trend (but no seasonality), series with seasonality (but no trend), and series with both trend and seasonality. They are denoted as `(N,N)', `(T,N)', `(N,S)' and `(T,S)', respectively. The trend and seasonality identification of a given time series is based on automatically fitting an ETS model using the function \code{ets()} of the \pkg{forecast} package in \proglang{R}.

\begin{figure}[ht]
  \centering
  \includegraphics[width=1\linewidth]{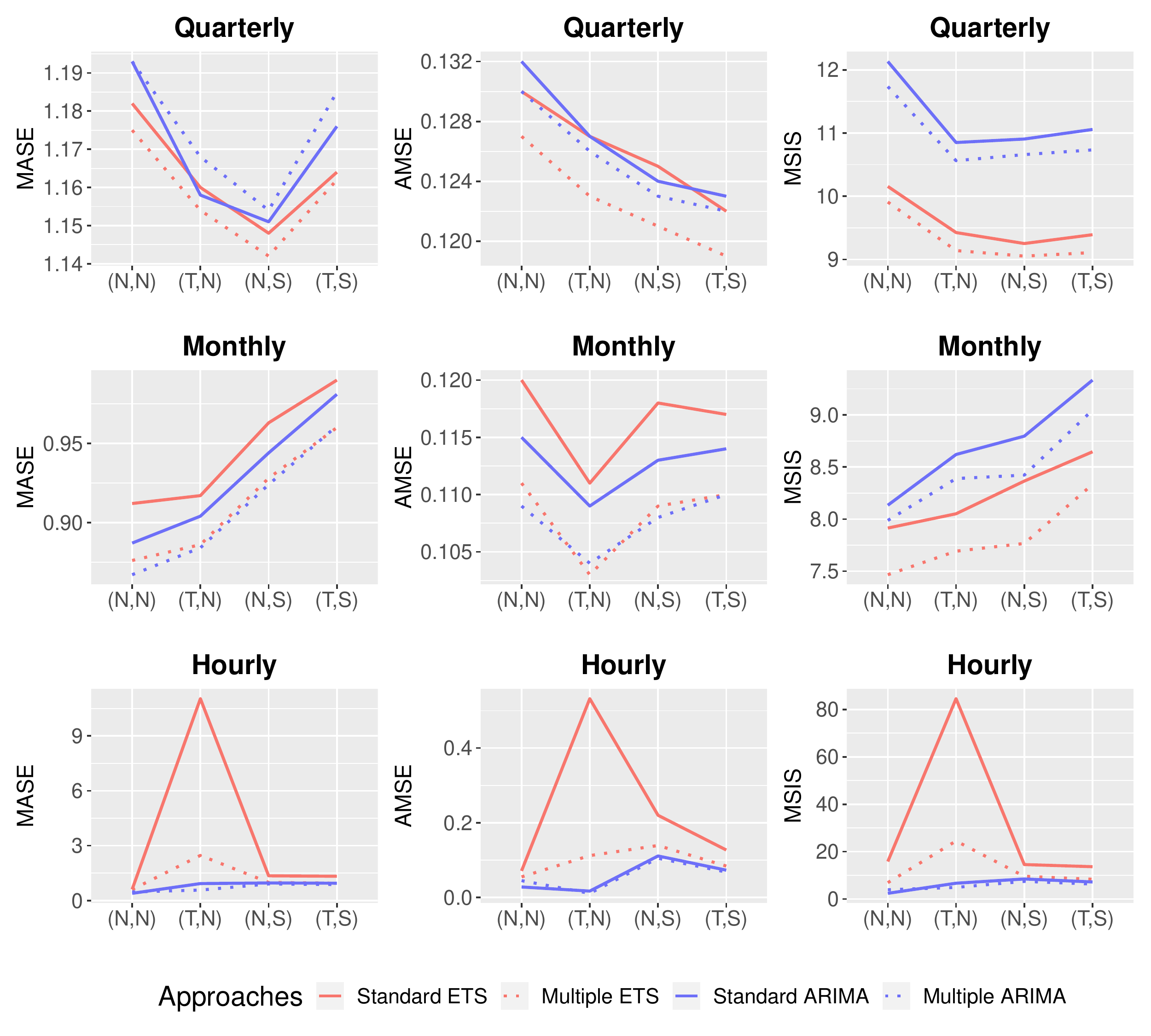}
  \\
 \caption{Caption: Benchmarking the performance of our proposed method against the benchmarks for series with different characteristics. `(N,N)', `(T,N)', `(N,S)' and `(T,S)' denote series with no trend nor seasonality, series with trend only, series with seasonality only, and series with both trend and seasonality, respectively.\\
 Figure~\ref{fig:features-performance}. Alt Text: A $3 \times 3$ plot showing nine line graphs to benchmarking the performance of our proposed method against the benchmarks for series with different characteristics.}
	\label{fig:features-performance}
\end{figure}

Figure \ref{fig:features-performance} visualises the forecasting performance of our proposed approach against the benchmarks with regard to the mean of MASE, AMSE, and MSIS values over the quarterly (first row), monthly (second row) and hourly (third row) series with different characteristics. We observe that:
\begin{itemize}
\tightlist
	\item For quarterly data and the ETS family of models, the proposed approach is consistently better than the benchmark (``Standard'') for each of the four categories considered.  Using ARIMA, the multiple approach does not improve forecasting in terms of point forecasting but still it provides better interval forecasts and less bias across all four categories. There are no noticeable differences in the scale of improvement or deterioration across the four categories, `(N,N)', `(T,N)', `(N,S)' and `(T,S)'. 
	\item For monthly data, ``Multiple'' always performs better than ``Standard'' for the data with different characteristics in terms of MASE, AMSE and MSIS. It appears that in terms of accuracy (MASE) and bias (AMSE), the improvements of ``Multiple'' over ``Standard'' are slightly larger when one or both of the time series patterns (trend and seasonality) are not picked up by ETS on the original series.
	\item For hourly data, ``Multiple'' almost always performs better than ``Standard'' for all the four data categories, except that for the `(N,N)' category where the multiple ARIMA method does not improve forecasting compared with the standard ARIMA. Moreover, focusing on ETS, the forecasts for the `(T,N)' category significantly benefit from using multiple sub-seasonal series.  
\end{itemize}

\FloatBarrier

\subsection{Analysis of the results with a focus on management of production systems and logistics}
As we mentioned in Section \ref{sec:data}, the empirical sets of data that we used for our empirical evaluation comprise of different data categories, which include demographic, industry, finance, economics, and others. In this subsection, we focus our attention on the results for the series within the ``industry'' category. \cite{PETROPOULOS2019251} also used the industry subset of the M3 competition data in a forecasting/inventory application context. After analysing the descriptions of these data, they concluded that the majority of the M3 industry series correspond to sales, shipments, or production. As such, focusing our investigation on the industry subsets of the M competitions, we aim to offer insights on how our proposed solution works towards improving the management of production systems and logistics. 

Table \ref{tab:industry} offers the summary results when the errors across all planning horizons (in this case, lead times) are averaged. We observe that our proposed solution that is based on multiple subseries and multiple models outperforms the state-of-the-art approaches (standard ETS and standard ARIMA). Using multiple ETS or multiple ARIMA results in better point forecast accuracy (MASE), lower bias (AMSE) and better estimation of the uncertainty (MSIS). The latter measure is relevant to inventory and logistics applications, as the accuracy of the prediction intervals can be associated with the inventory holding costs required but also the achieved service levels \citep{Svetunkov2018}.

\begin{table}[h]\small
\centering
\caption{The average forecasting performance of each criterion for each frequency and measure.}\vspace{0.25cm}
\begin{tabular}{ccccc|ccc}
\hline
& & \multicolumn{3}{c}{Quarterly} & \multicolumn{3}{c}{Monthly} \\
& & MASE & AMSE & MSIS & MASE & AMSE & MSIS \\
\hline
\multirow{2}{*}{ETS} & Standard & 1.143	& 0.104	& 9.236	& 0.983	& 0.113	& 8.260 \\
& Multiple & \textbf{1.138}	& \textbf{0.100}	& \textbf{8.943}	& \textbf{0.963}	& \textbf{0.107}	& \textbf{7.921} \\
\hline
\multirow{2}{*}{ARIMA} & Standard & 1.151	& 0.102	& 10.353	& 0.988	& 0.111	& 9.017 \\
& Multiple & 1.155	& \textbf{0.100}	& \textbf{9.952}	& \textbf{0.979}	& \textbf{0.108}	& \textbf{8.764}\\
\hline
\end{tabular}
\label{tab:industry}
\end{table}

\section{Applications to load forecasting}
\label{application}
\begin{figure}[ht]
  \centering
  \includegraphics[width=1\linewidth]{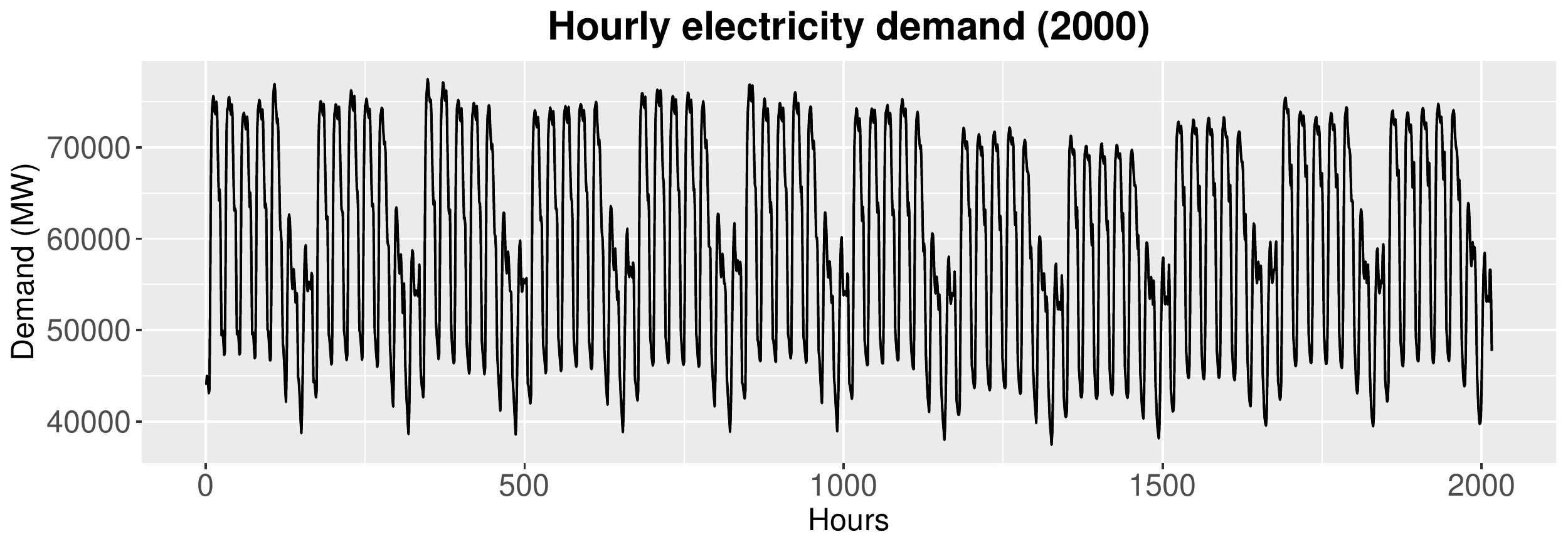}
    \includegraphics[width=1\linewidth]{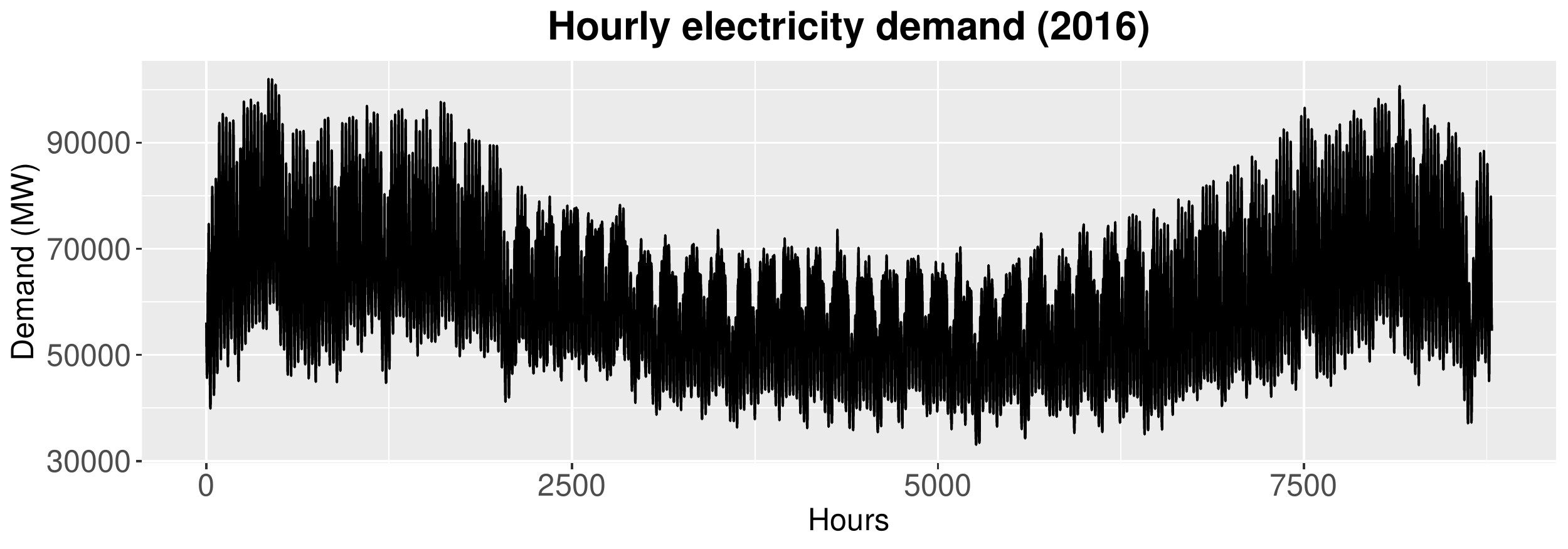}
  \\
 \caption{Caption: Top: Hourly electricity demand in England and Wales from 5 June 2000 to 27 August 2000. Bottom: Hourly electricity demand in England and Wales, 1 January 2016 to 31 December 2016.\\
	Figure~\ref{fig:2000-2016-elect}. Alt Text: Two panels of line graphs, with the top one showing the hourly electricity demand in England and Wales from 5 June 2000 to 27 August 2000, and the bottom one showing the hourly electricity demand in England and Wales, 1 January 2016 to 31 December 2016.}
	\label{fig:2000-2016-elect}
\end{figure}

In this section, we aim to further validate the good performance of our approach when complex seasonal patterns exist.
As a case study, we focus on two data sets of electricity load, which were also used in previous studies \citep{taylor2003short,rendon2019structural}.
The first data set is the hourly electricity demand in England and Wales from Monday 5 June 2000 to Sunday 27 August 2000~\citep{taylor2003short}, as depicted in the top panel of Figure~\ref{fig:2000-2016-elect}.  The second one is the hourly electricity demand in England and Wales from 1 January 2016 to 31 December 2016~\citep{rendon2019structural}, as shown in the bottom panel of Figure~\ref{fig:2000-2016-elect}. 

We focus on the hour within day and day within week patterns in this case study (the day within year pattern is not considered since the lengths of the two load data sets are both less than one year). As shown in Table~\ref{tab:subample-for-elect}, the standard that we apply is the Double-Seasonal Holt-Winters method \citep[DSHW,][]{taylor2003short} which considers two seasonal cycles (24 and 168, with respect to the hour within day and day within week patterns, respectively). DHSW is implemented using the \code{dshw()} function of the \proglang{R} \pkg{forecast} package. In the multiple approach, we consider 24 information levels. The subsampled series in the first level have single seasonality, while all the other levels exhibit double seasonality. Table~\ref{tab:subample-for-elect} presents our subsampling strategy for the electricity data with double seasonal patterns, together with the seasonal periods and the respective functions applied for each information level.

\begin{table}[ht]
  \caption{Standard and multiple approaches for load data with multiple seasonality implemented using the \proglang{R} package \pkg{forecast}.}
\label{tab:subample-for-elect}
\centering
\resizebox{\textwidth}{!}{
\begin{tabular}{p{0.15\columnwidth}cp{0.2\columnwidth}p{0.15\columnwidth}p{0.15\columnwidth}p{0.2\columnwidth}p{0.15\columnwidth}}
  \toprule
          Approach& Number of levels  & Subsamping & Seasonality & Period(s) &\proglang{R} function  \\
          \midrule
          Standard& 1 & None &Double &(24, 168) &\code{dshw()} \\
          \midrule
          \multirow{5}{5cm}{Multiple}&\multirow{5}{*}{24} &1h/day  &Single &7 &\code{hw()}        \\
          && 2h/day&Double &(2, 14) &\code{dshw()}\\
          && 3h/day&Double &(3, 21) &\code{dshw()}\\
          &&$\cdots$&\\
          && 24h/day&Double &(24, 168) &\code{dshw()}\\
          
  \bottomrule
  \end{tabular}}
\end{table}

\subsection{Forecasting hourly electricity demand in England and Wales from 5 June 2000 to 27 August 2000}

The top panel of Figure~\ref{fig:2000-2016-elect} shows the 12 weeks of hourly electricity demand in England and Wales from Monday 5 June 2000 to Sunday 27 August 2000. In line with \citet{taylor2003short}, the first eight weeks of data are used as the initial training data and the remaining four weeks are used to evaluate the post-sample forecasting performance of forecasts up to 24 hours ahead. As such, 1344 hourly observations are used for training and 672 for rolling-origin evaluation.

\begin{figure}[h!]
	\centering
	\includegraphics[width=1\linewidth]
	{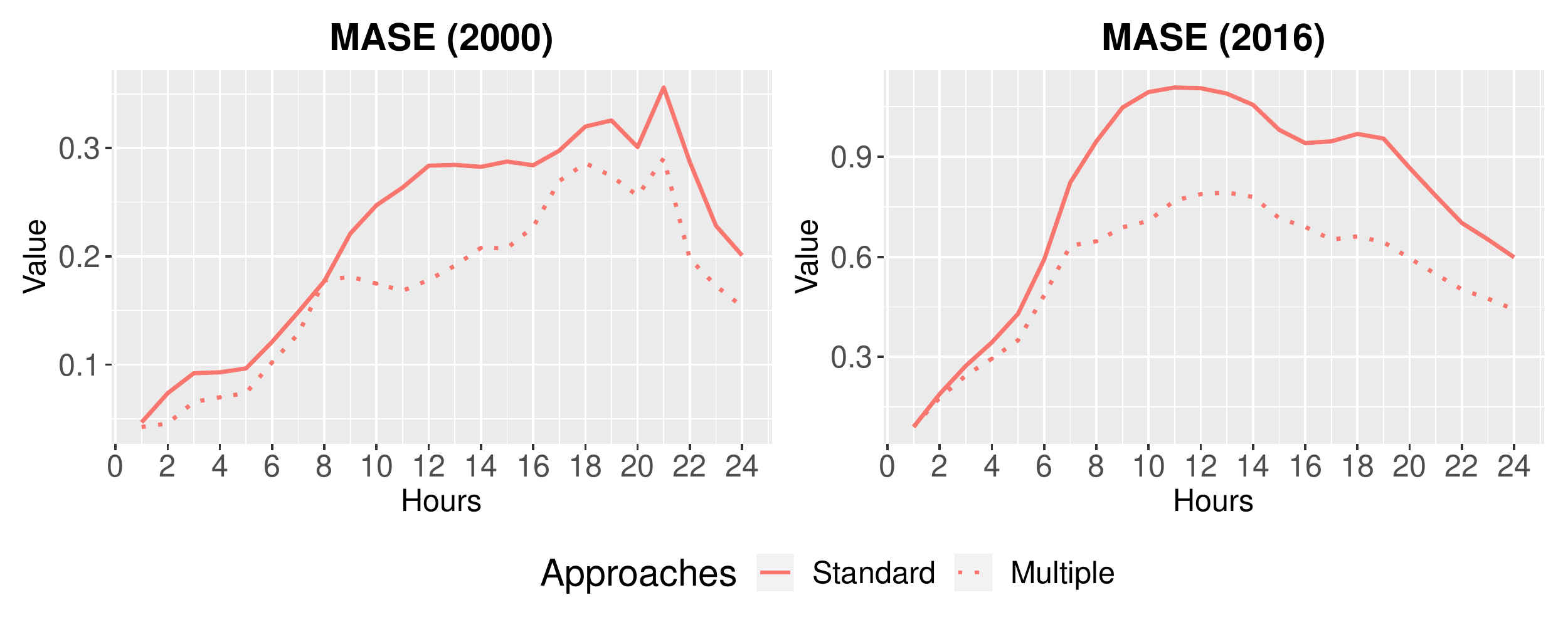}
	\caption{Caption: The forecasting performances of the standard and multiple approaches for all horizons over the hourly electricity demand in England and Wales for the year 2000 (from 5 June 2000 to 27 August 2000, left panel) and the year 2016 (from 1 January 2016 to 31 December 2016, right panel).\\
	Figure~\ref{fig:2000-2016-elect-mean}. Alt Text: Two panels of line graphs showing the forecasting performances of the standard and multiple approaches for all horizons over the hourly electricity demand in England and Wales for the year 2000 (from 5 June 2000 to 27 August 2000, left panel) and the year 2016 (from 1 January 2016 to 31 December 2016, right panel).}
	\label{fig:2000-2016-elect-mean}
\end{figure}

The left panel of Figure~\ref{fig:2000-2016-elect-mean} shows the forecasting accuracy for each forecasting horizon in terms of MASE using the standard and multiple approaches. We can observe significant benefits from using information from sub-seasonal patterns. Our approach performs better than the standard method over all horizons. On average, the multiple approach improves the forecasting performance by 22.12\% compared to the standard approach. Larger improvements are observed for horizons 10 to 16 hours ahead.

\subsection{Forecasting hourly electricity demand in England and Wales from 1 January 2016 to 31 December 2016}

The bottom panel of Figure~\ref{fig:2000-2016-elect} visualises the hourly electricity demand in England and Wales for the year 2016 (from 1 January 2016 to 31 December 2016). This is a much longer series compared to the one in the top panel of Figure~\ref{fig:2000-2016-elect}, and it will allow us to validate the usefulness of our approach for longer sequences.

To be consistent with \citet{rendon2019structural}, the series is split into a training period that consists of 35 weeks and a testing period of 17.3 weeks, with rolling forecasts up to 24 hours ahead being generated and evaluated. The right panel of Figure~\ref{fig:2000-2016-elect-mean} depicts the MASE values for the standard and multiple approaches. Compared with the standard approach, using information from multiple sub-seasonal levels improves the forecasting accuracy by 28.12\% on average. Significant improvements are observed for horizons 8 to 20 hours ahead.

\section{Discussion}
\label{discussion}

We are living in a big data era where large collections of time series are constantly generated. Forecasters often aim to choose the best model for their data automatically. However, accurate model selection requires professional knowledge and rich experience, making forecasting a difficult task. To mitigate the importance of model selection, we propose a novel seasonal time series forecasting method that constructs multiple series with diverse sub-seasonal patterns and subsequently extrapolates each new series separately. Finally, the forecasts of these subsampled series are averaged with equal weights.

In this paper, we propose the use of sub-seasonal patterns to forecast seasonal time series. Our approach makes it possible for forecasting methods to amplify time series patterns and hence improve forecasting performance. To make accurate and automatic extrapolations for these series individually, we use two widely-used statistical time series forecasting benchmarks, ETS and ARIMA, which are available in the \proglang{R} package \pkg{forecast}. Automatic and optimal model and parameter selection for these sub-seasonal series can ensure local optimal predictions.

We apply the proposed method on a large number of real-life series (the widely used data sets M1, M3, and M4) for empirical evaluation. We show that our approach performs better than the benchmarks in most horizons, whether in point or interval prediction for the monthly and hourly data sets, which indicates that our approach is more suitable for the time series with higher frequencies. On the other hand, the proposed approach does not improve as much over the standard benchmarks for the quarterly data sets. To compare the statistical significance of the accuracy improvements over the benchmarks, we carry out DM significance tests. The results further verify and strengthen the conclusion that our approach is more suitable for the time series with higher frequencies.

Our approach is also suitable for data with multiple seasonal cycles. We apply the proposed approach to the hourly electricity demand data that exhibit complex seasonal patterns to verify its effectiveness and stability. The empirical results show that our method indeed produces better and more robust point forecasts. 
Given that when ETS or ARIMA is used, different models might be involved at each information level when applying the multiple approach. Another implication of this case study is that our method also works well when only one model (DSHW) is used, highlighting the effect of subsampling. 

Why does the proposed approach work? We construct multiple time series consisting of one or several adjacent seasons, amplifying the sub-seasonal patterns and complex components of the original series. Furthermore, by extrapolating each new series separately and forecast combination, we effectively mitigate the importance of selecting a single model for (the original) series, making it possible to offer diverse and reliable forecasts. Our approach can significantly improve the performance of ETS and ARIMA for high frequency data through constructing multiple sub-seasonal series. More importantly, due to its simplicity and transparency, our method can be transferred in different contexts and works with other families of models.

Apart from the research implications, our approach also has clear implications for practice, particularly for production and logistics. Understanding the demand patterns and efficiently producing sales forecasts will not only lead to the maximisation of the customer service levels but also the minimisation of costs related to the inventory and logistics \citep{Doganis2008,wang2016select,Mircetic2021}. Sales patterns of retailers' stock keeping units are often recorded in high frequencies (daily) usually display strong seasonal patterns. Our approach can help in improving the forecasts of such historical sales, allowing for more informed decisions not only at an inventory but also a logistics (transportation from distribution centres to stores) and production level.

Our approach \textbf{fo}recasting with \textbf{s}ub-seasonal \textbf{s}eries (\pkg{foss}) is implemented as an \proglang{R} package. Our code is open-source and publicly available at \url{https://github.com/lixixibj/foss}.

\section{Conclusions}\label{conclusion}

To mitigate the importance of model selection, we propose a novel forecasting method that constructs new series containing the observations of only one or a few adjacent seasons of the original time series. Each subsampled time series is used to make forecasts for the corresponding seasons in the future. Finally, the forecasts produced for each horizon are combined with equal weights. The main advantage of our proposed method is that it makes full use of the sub-seasonal patterns that may be available within a time series. 

In our empirical experiments, the proposed approach yielded better forecasting performance than the benchmark methods for the widely-used forecasting competition data sets M1, M3, and M4. Our approach is particularly robust and stable for the data sets with higher frequencies, especially those with a trend or seasonal pattern. We also applied our approach to data with multiple seasonal cycles and showed its effectiveness in improving forecasting accuracy in the context of electricity demand.

In this paper, we focused our attention on creating sub-seasonal series using adjacent periods from the original seasonal series. An alternative would be to consider even more sub-seasonal series that are not necessarily limited to ones with adjacent periods, thus creating a larger pool of series and their accompanying forecasts. Averaging across such series could help us further reduce the variance of the forecasts even, possibly with a positive effect on the forecasting performance. Moreover, in this work, we set equal weights for combining the forecasts from the various sub-seasonal series. A potential avenue for future research would be investigating the performance of optimal (unequal) combination weights.
\section*{Data Availability Statement}
The M1 and M3 data sets that support the findings of this study are publicly available in the \pkg{Mcomp} \proglang{R} package~\citep{Mcomp}, and the M4 data set in the \pkg{M4comp2018} \proglang{R} package~\citep{montero2018m4comp2018}. Both the two data sets of electricity load for the application studies in Section~\ref{application} are publicly available at National Grid\footnote{\url{https://demandforecast.nationalgrid.com/efs_demand_forecast/faces/DataExplorer}}.

\section*{Acknowledgements}

Yanfei Kang is supported by the National Key Research and Development Program
(No. 2019YFB1404600) and the National Natural Science Foundation of China
(Nos. 72171011 and 72021001). This research was supported by the high-performance computing (HPC) resources at Beihang University. Fotios Petropoulos thanks Ivan Svetunkov for his feedback in the early-stages of the development of the algorithm proposed in this paper.

\bibliographystyle{agsm}
\bibliography{references}

\end{document}